\documentclass[twocolumn]{aastex631}

\shorttitle{Evolution of Warps}
\shortauthors{Brady et al.}

\setwatermarkfontsize{100pt}

\usepackage{xspace}
\usepackage{float}
\usepackage{soul}
\usepackage[version=4]{mhchem}
\usepackage[caption=false]{subfig}

\newcommand{\Me}{\ensuremath{M_{\oplus}}\xspace}

\newcommand{\Rsun}{\ensuremath{R_{\odot}}\xspace }
\newcommand{\Msun}{\ensuremath{M_{\odot}}\xspace}

\def\deg{\ensuremath{^{\circ}}}

\graphicspath{{./}{Figures/}}

\begin{document}

\title{Long-term Evolution of Warps in Debris Disks -- Application to the Gyr-old system HD 202628}




\correspondingauthor{Madison Brady}
\email{mtbrady@uchicago.edu}

\author[0000-0003-2404-2427]{Madison~T.~ Brady}
\affiliation{Department of Astronomy and Astrophysics, University of Chicago, Chicago, IL, 60637, USA}

\author[0000-0001-6403-841X]{Virginie Faramaz-Gorka}
\affiliation{Department of Astronomy and Steward Observatory, University of Arizona, 933 N Cherry Ave., Tucson, AZ 85721-0065, USA}

\author[0000-0001-5966-837X]{Geoffrey Bryden}
\affiliation{Jet Propulsion Laboratory, California Institute of Technology, 4800 Oak Grove Dr., Pasadena, CA 91109, USA}

\author[0000-0002-2314-7289]{Steve Ertel}
\affiliation{Department of Astronomy and Steward Observatory, University of Arizona, 933 N Cherry Ave., Tucson, AZ 85721-0065, USA}
\affiliation{Large Binocular Telescope Observatory, University of Arizona, 933 N Cherry Ave., Tucson, AZ 85721-0065, USA}




\begin{abstract}
We present the results of N-body simulations meant to reproduce the long-term effects of mutually inclined exoplanets on debris disks, using the HD 202628 system as a proxy.
HD 202628 is a Gyr-old solar-type star that possesses a directly observable, narrow debris ring with a clearly defined inner edge and non-zero eccentricity, hinting at the existence of a sculpting exoplanet. The eccentric nature of the disk leads us to examine the effect on it over Gyr timescales from an eccentric and inclined planet, placed on its orbit through scattering processes.
We find that, in systems with dynamical timescales akin to that of HD 202628, a planetary companion is capable of completely tilting the debris disk.  This tilt is preserved over the Gyr age of the system.  Simulated observations of our models show that an exoplanet around HD 202628 with an inclination misalignment $\gtrsim\,10\degr$ would cause the disk to be observably diffuse and broad, which is inconsistent with ALMA observations.

With these observations, we conclude that if there is an exoplanet shaping this disk, it likely had a mutual inclination of less than $5^{\circ}$ with the primordial disk. Conclusions of this work can be either applied to debris disks appearing as narrow rings (e.g., Fomalhaut, HR 4796), or to disks that are vertically thick at ALMA wavelengths (e.g., HD 110058).
\end{abstract}




\section{Introduction} \label{sec:intro}
 While there exist several prolific methods to determine the existence of exoplanets indirectly, such as Doppler spectroscopy, transit photometry, and astrometry, each of these techniques possesses various selection effects that make them only useful for measuring particular types of systems. More specifically, given the time baselines -- less than 30 years since the discovery of 51 Peg b \citep{Mayor1995} -- those methods are biased towards objects on relatively short, (sub-)Jovian period orbits.  The majority of known exoplanets have orbital periods less than 1000 days (see, e.g., the NASA Exoplanet Archive\footnote{https://exoplanetarchive.ipac.caltech.edu/index.html}).  However, direct imaging offers us the opportunity to study exoplanets on longer period orbits. Yet these exoplanets are still very difficult to detect directly. Direct imaging is itself biased towards planets that are still young -- typically a few tens of Myr old or younger -- and massive enough to produce an advantageous contrast with their host star \citep[see, e.g., the review by][]{Currie2022}.
Consequently, current ground-based instruments struggle to access sub-Jovian mass planets. 
This instrumental bias may explain the fact that the corresponding exoplanet parameter region -- super-Jovian periods \& sub-Jovian masses -- is underpopulated if not a desert, although it is still unclear from a theoretical point of view whether these exoplanets can exist \citep[see, e.g., the discussion by][and references therein]{Marino2018}.

At the dawn of the JWST era, another indirect method has already proven and goes on proving useful in this endeavour: the detailed examination of gravitational patterns in debris disks. Debris disks are the leftovers of planetary-system formation, and include populations of solids ranging from km-sized planetesimals down to micron sized dust grains that are in a collisional cascade \citep[see, e.g., the reviews of][]{Wyatt2008,Krivov2010}. Obvious examples close to us are the Solar System's Main Asteroid and Kuiper belts. And just as planets of the Solar System dynamically shape these debris belts, exoplanets will shape debris disks, leaving trace of their presence even when they escape classical detection methods. 
The study of these gravitationally induced structures can yield valuable information about the hidden planetary components of a system, and gives us access to this little known parameter space of exoplanet mass and period. Using this method, sub-Jovian mass planets at tens of AU have been inferred in systems such as HR 4796 \citep{Milli2017}, HD 107146 \citep{Marino2018}, HD 92945 \citep{Marino2019}, HD 206893 \citep{Marino2020}, AU Mic \citep{Daley2019}, and HD 15115 \citep{MacGregor2019}. 

In addition, most interestingly, this method can also gives us access to orbital characteristics that are otherwise difficult to obtain, such as eccentricity and the planet's mutual inclination with the disk \citep[see, e.g., the review of][]{Hughes2018}. Naturally, this method requires having access to resolved images of debris disks in the first place. As collisional evolution leads them to get fainter with age \citep{Wyatt2008}, there's also another important region of parameter space that is difficult to access: ages comparable to that of the Solar System. Consequently, opportunities to witness the long-term evolution of gravitational patterns are rare. 

\begin{figure}
\centering 
\includegraphics[width=1.\linewidth]{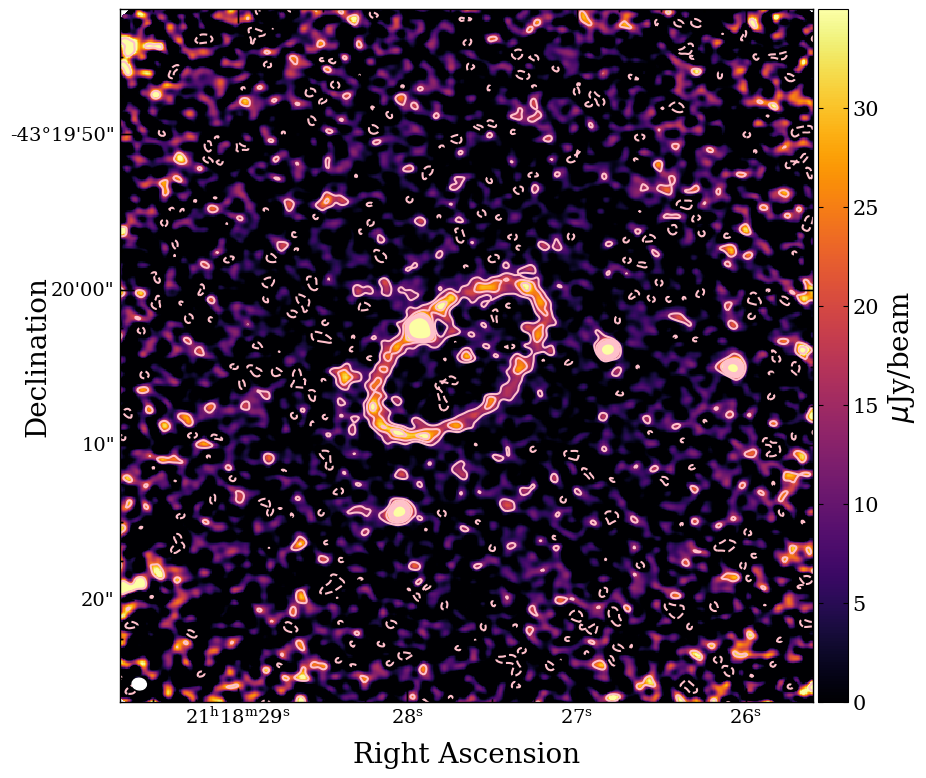}%
\caption{Clean ALMA image of the debris disk of HD 202628 \citep{Faramaz2019} at $\lambda\,=\,$1.3\,mm.  
Contours show the $\pm 2, 4, 6,...\,\sigma$ significance levels, with
$\sigma=5.2\,\mu\mathrm{Jy.beam}^{-1}$. The synthesized beam, shown on the lower left side of
the image, has dimensions $0\arcsec.92 \times 0\arcsec.75$, with position angle $83^\circ$.
}
    \label{fig:HD202628_ALMA}
\end{figure}

An excellent example of such an opportunity is the debris disk that revolves around HD 202628 (shown in Figure~\ref{fig:HD202628_ALMA}, and whose parameters are listed in Table \ref{tab:star}.  This star possesses solar-like properties and is Gyr-old, and the system's proximity and large radial extent have afforded the disk an angular separation -- $\sim 6\arcsec$ -- such that it can clearly be resolved from its host star. 

\begin{deluxetable*}{ccc}[tbp]
\tablecaption{Parameters of the HD 202628 system.} \label{tab:star}
\tablewidth{0pt}
\tablehead{
\colhead{Parameter} & \colhead{Value} & \colhead{Reference}
}
\startdata
\multicolumn{3}{c}{{\bf Stellar parameters}} \\[2pt]
\hline
Spectral Type                  & G1.5V               & \cite{Gray2006} \\[2pt]
Effective Temperature $T_\mathrm{eff}$ (K)         & $5833   \pm 6$      & \cite{Spina2018}\\[2pt]
Metallicity {[Fe/H]} (dex)          & $0.003  \pm 0.004$  & \cite{Spina2018} \\[2pt]
Surface Gravity $\log(g)$       & $4.51   \pm 0.01$  & \cite{Spina2018} \\[2pt]
Distance $d$ (pc)         & $23.794 \pm 0.012$  & \cite{GaiaEDR3} \\[2pt]
Age (Gyr)             & $1.1    \pm 0.4$    & \cite{Faramaz2019} \\[2pt]
Mass $M_{\star}$ (\Msun)& $1.068  \pm 0.038$  & \cite{Faramaz2019} \\[2pt]
Radius $R_{\star}$ (\Rsun).       & $0.951  \pm 0.013$  & \cite{Faramaz2019} \\[2pt]
\hline
\multicolumn{3}{c}{{\bf Disk Parameters}} \\[2pt]
\hline
Eccentricity $e_d$                 & $0.09^{+0.02}_{-0.01}$      & \cite{Faramaz2019} \\[2pt]
Inner Semimajor Axis $a_\mathrm{inner}$ (AU)   & $143.1 \pm 1.7$             & \cite{Faramaz2019} \\[2pt]
Outer Semimajor Axis $a_\mathrm{outer}$ (AU)   & $165.5 \pm 1.4$             & \cite{Faramaz2019} \\[2pt]
mm Dust Mass $M_{d}$ (\Me)         & $(1.36 \pm 0.06) \times 10^{-2}$ & \cite{Faramaz2019} \\[2pt]
Inclination $i_d$ ($\deg$) & $57.4 \pm 0.4$              & \cite{Faramaz2019} \\[2pt]
Position Angle $\Omega_d$ ($\deg$)          & $-50.4^{+0.4}_{-0.5}$       & \cite{Faramaz2019} \\[2pt]
\enddata
\end{deluxetable*}

Observations with the \textit{Hubble Space Telescope} (HST) reported by \citet{Krist2012} show that the debris disk has a sharp inner edge and that the star is offset from the disk center of symmetry, indicating a disk with intrinsic eccentricity of $\sim 0.1$. Further observations with the Atacama Large Millimeter Array (ALMA) confirmed this view and yielded further, more precise information about the density structure of the disk and its extent \citep{Faramaz2019}. The disk eccentricity and sharp inner edge suggest it is dynamically shaped by an eccentric exoplanet orbiting interior to the disk, at a separation $\gtrsim 100$\,AU from the central star.  As described in \cite{Kennedy2020}, the ring is narrower than expected given its semi-major axis and eccentricity, which may indicate that the forcing planet may have excited the  disk's eccentricity before the gas-rich protoplanetary disk dispersed. A planet-free formation scenario of eccentric rings is also possible \citep{Lyra2013}, but current constraints on the system cannot rule out the presence of a massive planetary companion \citep{Rodigas2014}.

There is growing evidence for exoplanets in this separation regime. Their presence is usually attributed to three classes of phenomena, in situ formation, planetary migration and planet-planet scattering. However, in situ formation is still actively debated, and migration processes are expected to damp orbital eccentricities \citep{Ward1997,Masset2003,Ida2008,Crida2009}. Thus the existence of long-period planets on eccentric orbits such as the one posited around HD 202628 is preferentially attributed to a planet-planet scattering event \citep[see the review of][and references therein]{Raymond2022}. The narrow width of the debris disk could also be explained by past scattering events, though scattering events can both narrow and broaden disks \citep{Rodet2022}.

As developed further in Section~\ref{ssec:planet_params} of this paper, when examining the outcome of simulations of such scattering events by \citet{Carrera2019}\footnote{The full set of resulting orbital elements -- including the orbital inclinations -- was obtained via private communication with the authors.}, we observed that these can cause planets to develop mutually inclined orbits in addition to excited eccentricities.  Their scattering events were clearly capable of affecting the entire set of orbital parameters, and could excite the planets' inclinations even without dramatically exciting the eccentricity or kicking the planet out of the system.  These inclined companions can strongly influence the shape of a debris disk \citep[as seen in simulations from, e.g.,][]{Mouillet1997, Pearce2014, Kennedy2012}.  If a planet was able to scatter out to a large orbital period and heightened eccentricity/inclination while ejecting any nearby companions, the disk's shape could be used to infer information about the stirring planet.

As an example, models of the warps observed in the $\beta$ Pictoris disk have been used to help constrain the inclination of its directly imaged stirring planet, as well as the presence of any additional planets and even some elements of the disk dynamics \citep{Mouillet1997, Dawson2011, Nesvold2015}.  Given the potential for an inclined planet to affect the observational signature of a nearby disk, it is important to study the impact of a mutually inclined exoplanet on the HD 202628 disk over its lifetime.

\cite{Kennedy2020} analytically modeled the HD 202628 disk and found only very weak constraints on the forced inclination of the disk, showing it was below 30\degr.  In this paper, we perform a full set of N-body simulations over 1 Gyr to explore the possibility of inclined exoplanets in the HD 202628 system and to provide more precise constraints on their parameters.

The programs we used and the assumptions we made with regards to our N-body simulations are detailed in Section \ref{sec:sim}.  The results of our simulations are presented in Section \ref{sec:results}, and we discuss their implications in Section \ref{sec:disc}. Finally, we present our conclusions in Section \ref{sec:conclu}.

\section{Numerical Simulations} \label{sec:sim}
Our goal is to explore the gravitational impact of a planet mutually inclined with the debris disk of HD 202628 on Gyr timescales. In this section, we present the assumptions we made and the subsequent N-body simulations that we carried out in order to compare them  with the ALMA observations of this debris disk.  We will also describe the setup for these simulations.  The summary of the planetary parameters explored in our N-body simulations is presented in Table~\ref{tab:planet}.

\subsection{General Setup and Assumptions}
\label{ssec:nbody}

For the sake of modelling this debris disk, we considered the system to be a set of numerous independent 3-body problems consisting of a central host star, the planetary perturber, and a massless test particle. We integrated the dynamical evolution of those test particles over 1\,Gyr using the symplectic code SWIFT-RMVS of \cite{Levison1994}.  We chose to run the simulations over 1\,Gyr as it is roughly the same order of magnitude as the ages predicted for the system through various methods \citep[see][and the references therein]{Faramaz2019}.  The initial time, $t\,=\,0$, represents the time at which the planet (which was initially on a circular orbit that was coplanar with the disk) reaches its final mutual inclination with the disk and eccentricity as the result of a scattering process with some other object. This second object is not accounted for in our simulations, as there is no evidence to motivate modelling a second planet in the system.  It is reasonable to assume that this secondary object has been scattered out of the system, as the \citet{Carrera2019} simulations show that the scattering interactions that produce long-period planets also frequently eject other planets from the system.

The integration timestep was set to be $\frac{1}{20}$ the smallest orbital period involved, and we recorded snapshots of the simulation every 10\,Myr.
In order to account for particles leaving the system, the simulations treated particles with an semimajor axis less than 0.05\,AU as having collided with the host star and particles with a semimajor axis greater than 1000\,AU as having been scattered out of the system.


Our simulations do not include stellar radiation effects, that is, radiation pressure and Poynting-Robertson drag. These effects tend to make dust grains drift far from where they were collisionally produced. The result is that spatial structures induced by perturbing planets tend to be smeared in observations that trace small grains, which are significantly affected by these stellar radiation effects \citep[see, e.g., ][]{Thebault2007,Thebault2012,Ertel2012}. This is the case in scattered light or at mid-infrared wavelengths, which trace micron-sized dust grains. However, as the ALMA observations were taken at 1.3\,mm, we expect the observed thermal emission of the HD 202628 disk to be dominated by mm-sized grains \citep{Ertel2011}. These are little affected by radiation pressure or Poynting-Robertson drag and hence excellent tracers of gravitationally induced patterns. In addition, we expect these grains to be the product of a collisional cascade, where grains collisionally destroyed are constantly replaced, such that their population is considered to be at steady-state. Consequently, it is reasonable to expect that our simulations, which do not model particle collisions, are capable of reproducing the observational conditions.

In general, we do not expect it to be necessary to model the effect of particle collisions when considering the 1.3 mm ALMA images.  As discussed in \citet{Lohne2017}, dust collisons cause observable asymmetric halos with reduced pericenter glow and more tenuous apastrons,  but this effect is most prominent when studying very small grains.  Their simulations of dust collisions at $\lambda\,=\,1.2$\,mm demonstrate a degree of apocenter glow comparable to the glow of the collisionless models in \citet{Pan2016}, indicating that, for images at these wavelengths, the effects of collisions are likely minor.

\citet{Dong2020} showed that debris disks are typically the most influenced by the nearest planet, while their inclinations and warps are most heavily influenced by the most massive planet.  We assume that there is only one planet in the system at these large semi-major axes, and thus this single planet has to be responsible for both the eccentricity and inclination of the disk at the time of observation.  There is a possibility that there are multiple planets which influence the disk eccentricity and inclination separately, but we do not fully explore this case to reduce the complexity and computation time of our simulations.  It is reasonable to assume that the interactions that produced the long-period disk-shaping planet ejected one or more planets from the system, as the simulation results from \citet{Carrera2019} showed that planets were only able to achieve semi-major axes larger than 50 AU when one or more planets were ejected from the system. If these scattering and ejection events occurred over shorter timescales than the secular timescale of the disk, we would not expect any of these ejected planets to have any influence on the shape of the disk.  \cite{Carrera2019} found that that ejection timescales in their simulated systems typically ranged from around $10^4\,$--$\,10^6$\,yr, meaning that such rapid ejections are possible, especially for systems with low-mass planets.


%

\subsection{Disk Initial Conditions}
\label{ssec:ring_init}

Our simulations initially consist of a ring of 50,000 particles in orbit around a $1\,\Msun$ star, with initial eccentricities ranging from 0.00 -- 0.05 and inclinations from -3 -- 3\deg, in line with expectations for a cold debris disk and observations of the cold component of the Kuiper Belt by \citet{Petit2011}.  In general, we expect a low primordial eccentricity and inclination dispersion due to the action of the circumstellar gas during the protoplanetary phase.  \citet{Kennedy2020} hypothesized that the narrow width of the HD 202628 ring can be explained by a model in which the perturbing planet is able to excite the particle eccentricities before the gas disk dissipated.  In addition, \citet{Rodet2022} also showed that the damping of planetesimal eccentricity by disk gas or the slow growth of planetary eccentricity can also result in narrow debris rings. However, we are not interested in precisely modeling the disk's width so we do not model the early portion of the disk's lifetime in which these interactions would be important.

The longitude of periastron $\omega$, the longitude of ascending node $\Omega$, and the mean anomaly $M$ for each particle, were randomly drawn from a uniform distribution between 0 and 2$\pi$.

There is no way to determine the pre-planet disk width in the system HD 202628 given observations. Particles belonging to the planet's chaotic zone \citep[within $\sim 3.5$ Hill radii of the planet, ][]{Ida2000,Kirsh2009} are rapidly cleared from the system during our simulations, so our results are insensitive to their inclusion. Thus, our simulations were initialized with the final disk semimajor axis and width in order to limit the number of close encounters with the planet, as we found that these interactions drastically increased the computation time. Consequently, the disk's inner and outer edges -- $a_\mathrm{inner}$ and $a_\mathrm{outer}$, respectively -- were initialized at 143.1 and 165.5\,AU, as per observations from \cite{Faramaz2019}, with each test particle's semimajor axis being drawn from a uniform distribution in this range.  This simplifying gesture prevents us from making any more detailed inferences on the disk width- for a more detailed discussion of the width of the debris disk and possible causes, see \cite{Kennedy2020}.

\subsection{Planetary Parameters}
\label{ssec:planet_params}

\paragraph{Planet Inclination}
In order to explore the impact of an exoplanet on a mutually inclined disk, we first must determine which planet inclinations we can expect to see as a result of a planet-planet scattering event.

\cite{Carrera2019} examined the end result of gravitational scattering events amongst three $1\,M_\mathrm{Jup}$ planets on coplanar orbits from $\approx 3\,$--$\,7$\,AU.  While there was a strong positive correlation between the post-scattering semi-major axes and eccentricities of the planets in their sample, such a correlation was not obvious for the planet inclinations.  While the majority of the planets scattered out to semi-major axes beyond 100 AU in their simulations had eccentricities above 0.8, over 80\% of these post-scattering planets had inclinations below 25$^o$.  These planets thus seem to typically retain somewhat low inclinations even after ejecting other planets from the system.  While we cannot guarantee that a similar scattering interaction produced the HD 202628 system (we included planets with $M_p \neq 1\,M_\mathrm{Jup}$ in our simulations), \cite{Carrera2019} showed that their modeled scattering interactions produced planets with eccentricities consistent with the observed eccentricity distribution of high-eccentricity exoplanets. Thus, their results provide a reasonable upper bound on the energetic nature of any hypothetical orbital scattering events in the HD 202628 system.

Assuming that a typical planet begins its life orbiting in the same plane as the protoplanetary disk and undergoes scattering interactions similar to or less energetic than those in \citep{Carrera2019}, we restrict our simulations to those with initial planet mutual inclinations of $I_p \leq 25\deg$, examining $I_p = 0\deg,~5\deg,~10\deg,~15\deg,~20\deg,$ and $25\deg$.  This choice of an upper limit is in line with the observational constraints from \cite{Kennedy2020} of the HD 202628 disk.

\paragraph{Mass}
The more massive a planet is, the wider the area over which it exerts gravitational influence. Thus, a low-mass planet orbiting closely to a debris disk edge can produce a debris disk with an inner edge at the same orbital separation as a more massive planet with a shorter-period orbit. This degeneracy makes it impossible to pinpoint the precise mass of the perturbing planet based on the location of the disk inner edge alone.  We thus chose to model a grid of planet masses, with $M_p = 0.1,~0.5,~1,~5,$ and $10\,M_\mathrm{Jup}$.
We selected the upper bound by taking into account the simulations from \cite{Rodigas2014}, which estimated the maximum mass of the HD 202628 companion by studying the disk width in HST data, finding $M_p < 15.4 \pm 5.5\, M_\mathrm{Jup}$. Regarding the lower bound, we based our choice of $M_p = 0.1\,M_\mathrm{Jup}$ upon \cite{Faramaz2019}, which argues smaller planets would be unable to stir the debris disk eccentricity on timescales comparable to the age of the system.


Our usage of massless test particles means that we do not take into account either the self-gravity of the disk or the gravitational influence of the disk on the planet.  In general, these phenomena are significant only when the mass of the disk is comparable to (or greater than) the mass of the planet. It is thus important to compare the disk mass to the modeled planet masses for the purposes of discussing the validity of our chosen techniques. While \citet{Faramaz2019} found that the mass of mm dust in the HD 202628 disk was around $10^{-2}\,\mathrm{M}_\oplus$ (which is far less than our smallest simulated planet mass), we lack constraints on the total mass of the disk.  However, we can compare the HD 202628 disk to that around Fomalhaut, which is both younger \citep[at an age of around 400 Myr, see][]{Mamajek2012} and brighter in 1.3 mm ALMA emission \citep[even when considering their relative distances, see]{MacGregor2017}, and is thus expected to be more massive.  Fomalhaut has been extensively studied, and its disk has been found to have a mass of $\sim\,3-30\,\mathrm{M_\oplus}$ \citep{Wyatt2002, Chiang2009}.  If the HD 202628 disk is less massive than that around Fomalhaut, it is valid to assume that the effects of a massive disk would be negligible given a minimum modeled planet mass of around 30\,M$_\oplus$.  A similar argument was used in \citet{Faramaz2014} when modeling the debris disk of $\zeta^2$ Reticuli.

\paragraph{Semimajor axis}
To determine where the planetary orbits should be placed in the system, we considered the reasoning detailed in \cite{Pearce2014}. It describes $Q_\mathrm{edge}$, the apastron of the inner edge of the disk -- reasonably approximated as being the semimajor axis of the inner edge -- as a function of the planet's apastron $Q_{p, a}$:
\begin{equation}
\label{eqn:qedge}
Q_\mathrm{edge} \approx Q_{p, a} + 5 R_{H, Q} \quad,
\end{equation}
where $R_{H, Q}$ is the planet's Hill radius at apastron, and is itself a function of the planet's semimajor axis and mass:
\begin{equation}
\label{eqn:hill}
R_{H, Q} \approx a_p (1 + e_p) \bigg( \frac{M_p}{(3-e_p) M_{*}} \bigg) ^{1/3} \quad.
\end{equation}

Using these equations and the disk inner edge of $Q_\mathrm{edge} = 143.1$\,AU \citep[from ALMA observations;][]{Faramaz2019}, we determined the planetary semimajor axes $a_p$ necessary to generate a ring with the observed inner edge given each planet's mass.

\paragraph{Eccentricity}

To account for the eccentricity of the observed disk, we chose to model the exoplanet as possessing a fixed eccentricity over the course of our Gyr simulation.  The system is old enough that the gas surface density is low, making gas-induced eccentricity damping unlikely \citep[see the equations in][]{Tanaka2004}.  Additionally, as described in previous sections, the planet is likely massive enough that it is unlikely for the disk to have any meaningful gravitational influence upon it.  Using Laplace-Lagrange secular theory to model the three-body problem and assuming $e$ is small, we selected an eccentricity for the exoplanet based off of the following equation from \cite{Wyatt1999}:

 \begin{equation}
\label{eqn:ef}
e_f \simeq \frac{b^2_{3/2}\Big( \frac{a_p}{a} \Big)}{b^1_{3/2}\Big( \frac{a_p}{a} \Big)} e_p,
\end{equation}

where $e_f$ is the eccentricity forced onto a disk of semimajor axis $a$ by an exoplanet with an eccentricity $e_p$ and semimajor axis of $a_p$, and the $b$ terms are Laplace coefficients.  This equation is valid when ${a_p}/{a}\,<\,1$.  In the limit of small ${a_p}/{a}$, we can make leading-order approximations to the Laplace coefficients \citep[as is done in][]{Mustill2009} and find

\begin{equation}
\label{eqn:ef2}
e_f \simeq \frac{5}{4} \Big( \frac{a_p}{a} \Big) e_p \quad,
\end{equation}

As we are modelling the planet as having an orbit interior to the disk, this equation can thus be used.  However, ${a_p}/{a}$ is not always small in our simulations (especially when the planet is of a very low mass), so this approximation may not always hold.

Using numerical integration to evaluate the Laplace coefficients, we find that, to create a disk with an eccentricity of about 0.09 \citep[reflective of HD 202628's observed eccentricity, see][]{Faramaz2019}, the eccentricities of our modeled exoplanets should be between roughly 0.11 and 0.15, with the exact eccentricity value dependent on the mass and semi-major axis of the exoplanet.  This range of eccentricities is comparable to the 10-20\% observational error on the disk eccentricity measured by \citep{Faramaz2019}, as well as the difference between the eccentricity measured by \cite{Faramaz2019} and \cite{Kennedy2020} ($e_f\,=\,0.09$ and 0.12, respectively).  It is thus acceptable to model the planet in each simulation with the same orbital eccentricity, which we fix at $e_p\,=\,0.10$ in order to roughly reproduce the observational conditions.  As our intention is to explore the long-term effects of the planet on the disk as opposed to modelling the exact eccentricity of HD 202628, slight deviations in simulated eccentricity from the observations are unlikely to influence our results, especially given the uncertainties.

Equation~\ref{eqn:ef2} also gives us insight into the expected surface brightness distribution of the disk.  The apocenter glow effect, described in \citet{Pan2016}, describes a situation in which the disk appears brighter at its apocenter at long wavelengths.  This is due to a pile-up of particles at the disk apocenter because of their lower relative velocities.  However, as noted in \citet{Marino2019} and \citet{Lynch2022}, the particles at apocenter are more spread out radially, so a well-resolved disk with a constant eccentricity would have no obvious apocenter glow effect.  The apocenter glow can still be observed, however, if the disk width is poorly resolved or if the disk eccentricity decreases with distance.  Equation~\ref{eqn:ef2} shows that, for a fixed planet eccentricity, we expect the disk particle eccentricity to decrease with semi-major axis, though given the narrow width of the disk it is difficult to conclude whether or not the decrease in eccentricity is extreme enough to produce a glow effect given the image resolution.  \citet{Faramaz2019} did find tentative evidence of apocenter glow in the HD 202628 disk, but they noted the detection was inconclusive.

\paragraph{Orientation}

We set both the planet's longitude of ascending node $\Omega_p$ and longitude of periastron $\omega_p$ to $0\deg$. This means that the periastron and ascending node share the same direction. This is not expected to be necessarily true and any combination of ($\Omega_p$,$\omega_p$) could occur as a result of a scattering event. This is, however, not expected to affect our findings on the capacity of an inclined planet to tilt a debris disk, since the secular effects of inclination and eccentricity are independent and decoupled according to the Laplace-Lagrange theory \citep{Murray1999}.  A more expansive exploration of $\Omega_p$ and $\omega_p$ would potentially be necessary if we found that the disk was consistent with a highly inclined planet.

\begin{deluxetable}{cccc}[htbp]
\tablecaption{Modeled Planet Parameters. \label{tab:planet}}
\tablewidth{0pt}
\tablehead{
\colhead{Mass} &
\colhead{Semimajor axis} &
\colhead{Eccentricity} &
\colhead{Initial Planet Inclination} \\
\colhead{(M$_\mathrm{Jup}$)} &
\colhead{$a_p$ (AU)} &
\colhead{$e_p$} &
\colhead{$i_p$ ($\deg$)}
}
\startdata
0.10          & 112.1                  & 0.1          & 5, 10, 15, 20, 25 \\ [2pt]
0.50          & 102.1                  & 0.1          & 5, 10, 15, 20, 25 \\ [2pt]
1.00          & 96.7                   & 0.1          & 5, 10, 15, 20, 25 \\ [2pt]
5.00          & 81.8                   & 0.1          & 5, 10, 15, 20, 25 \\ [2pt]
10.0          & 74.6                   & 0.1          & 5, 10, 15, 20, 25 \\ [2pt]
\enddata
\end{deluxetable}

\section{Results} \label{sec:results}

\subsection{Tilting the Debris Disk}
\label{ssec:tilt}

\begin{figure*}
\centering
\subfloat{%
  \includegraphics[width=.5\linewidth]{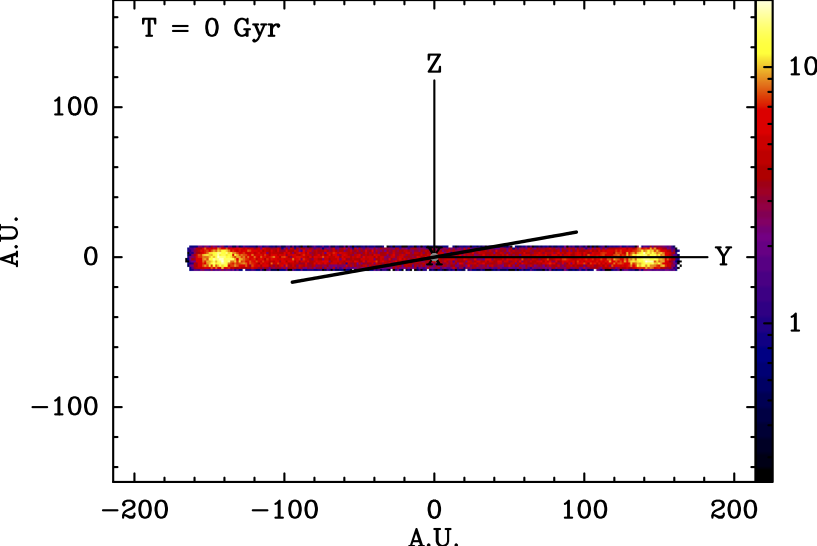}%
  \label{fig:sfig1_initial}%
}
\subfloat{%
  \includegraphics[width=.5\linewidth]{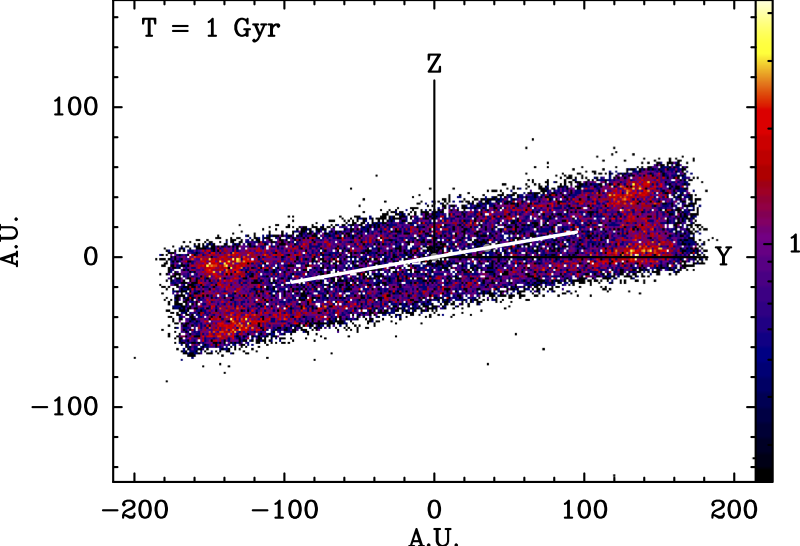}%
  \label{fig:sfig2_final}%
}
\caption{Edge-on density maps of the initial (left) and final (right) state of a debris disk perturbed by a planet of $M_p = 1\,M_j$ mutually inclined by $I_p = 10\deg$ relative to the initial disk.  A black line (left panel) and white line (right panel) show the edge-on orbit of the planet.  Note that the disk particles develop inclinations ranging from $0\deg$ to twice the planetary inclination.}
\label{fig:1m10i_disk}
\end{figure*}

We found that all of our simulated planets were capable of completely inclining the entire radial extent of their debris disks over the course of 1\,Gyr. Each disk had a range of final particle inclinations varying between 0 -- $2\,I_p$ degrees, with both proper and forced  inclinations of $I_p$ degrees.  The particle inclinations and $\Omega$ were related to one another as predicted by \citet{Wyatt1999hr4796}.  Figure~\ref{fig:1m10i_disk} shows the difference between the initial and final particle distribution for the $M_p = 1\,M_\mathrm{J}$ model in a side-on view, demonstrating how the disk orientation and shape change over time.

As discussed in, for example, \citet{Wyatt1999hr4796} and \citet{Pearce2014}, the timescale characterizing the onset of a particle on an inclined orbit is the secular timescale $t_{sec}$, as predicted by second-order theory by:
\begin{equation}
\label{eqn:tsec}
t_{sec} \approx 4 T_p \left( \frac{M_p}{M_{*}} \right) ^{-1} \alpha ^{-5/2} \bigg[ b^{(1)}_{3/2} \Big( \alpha \Big) \bigg] ^{-1} ,
\end{equation}
where $T_p$ is the planet's orbital period and $b^{(1)}_{3/2}(\alpha)$ is a Laplace coefficient as defined in \cite{Murray1999}, and $\alpha$ is the ratio of the semimajor axes of the particle $a$ and the planet $a_p$, such that this ratio is always inferior to 1. In that case, since $a_p<a$, we have $\alpha=a_p/a$.

From this expression, and as seen in our simulations, the time it takes to tilt a disk is completely independent of $I_p$. In addition, more massive planets tilt the disk more quickly, as we expect. Estimating $t_{sec}$ with a numerical calculation of $b^{(1)}_{3/2}(\alpha)$ shows that all of the planets modeled had 10$t_{sec}<$1\,Gyr, demonstrating that our results agree with analytic theory and \citet{Pearce2014}, which found that disk particles took on the order of a few to ten times the secular timescale to settle onto their final inclined orbits.

We can use this equation to revisit our assumption that the disk perturbing planet would be able to eject any nearby planets in the system before they could have any meaningful influence on the disk's shape.  $t_{sec}$ for the closest edge for the disk ranges from $0.5\,$--$\,6.5$\,Myr, with the largest simulated planet having the shortest timescale.  As \cite{Carrera2019} found that the first ejection/collision event in their scattering simulations typically happened between about 35\,kyr and 4.6\,Myr, it is possible for additional planets in the system to be ejected before they can perturb the disk.  However, a very massive planet ($M\,>\,1\,M_\mathrm{Jup}$) could have some influence on the disk before being kicked out unless it is removed from the system extremely quickly.

The precise dependency on the planet's semimajor axis is not obvious in Equation~\ref{eqn:tsec}, as part of the semi-major axis dependency is contained within the Laplace coefficient. This term can be expanded in terms of $\alpha$ at small values of $\alpha$, and to first order \citep[see Equation (6.68) of][]{Murray1999}, $b^{(1)}_{3/2}(\alpha) \approx 3\alpha$. This yields us the equation
\begin{equation}
\label{eqn:tsec2}
t_{sec} \approx \frac{4}{3} T_p \left( \frac{M_p}{M_{*}} \right) ^{-1} \alpha ^{-7/2} \qquad.
\end{equation}

At larger values of $\alpha=a_p/a$ the dependencies are more complex, but the general trends remain the same.  In general, the particles that are the closest to the planet, with the largest $a_p/a$, are forced on inclined orbits the fastest. Since the disk in our simulations is narrow and hence does not span a wide range of semimajor axes, this effect is not strikingly visible.

Tests done with larger initial disk widths ($a_{disk} = 143.1-1000$\,AU) showed that particles at larger values of $a$ develop excited inclinations more slowly. At intermediate times, this results in a system in which the particles nearer to the planet develop excited inclinations while more distant particles retain their low primordial inclinations, generating a disk that appears to have two sections that are mutually inclined relative to one another. In other words, we have a disk the exhibits a warp, and since the planet is interior to the disk, the warp will develop in the inner parts of the disk and expand outwards. This is exactly the type of situation that is seen in the system of $\beta$ Pictoris \citep{Mouillet1997}.

\begin{figure*}
\centering
\subfloat{%
  \includegraphics[width=.45\linewidth]{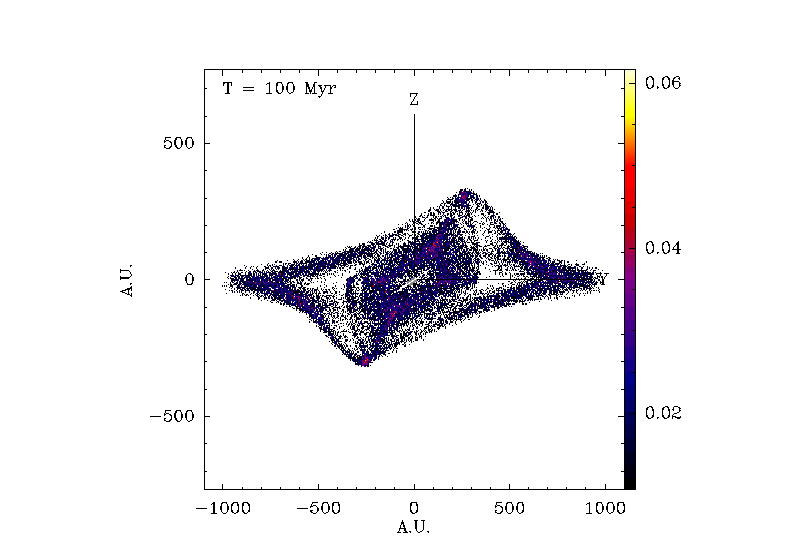}%
  \label{fig:sfig1_side}%
} \hspace{0.05\linewidth}
\subfloat{
\includegraphics[width=.42\linewidth]{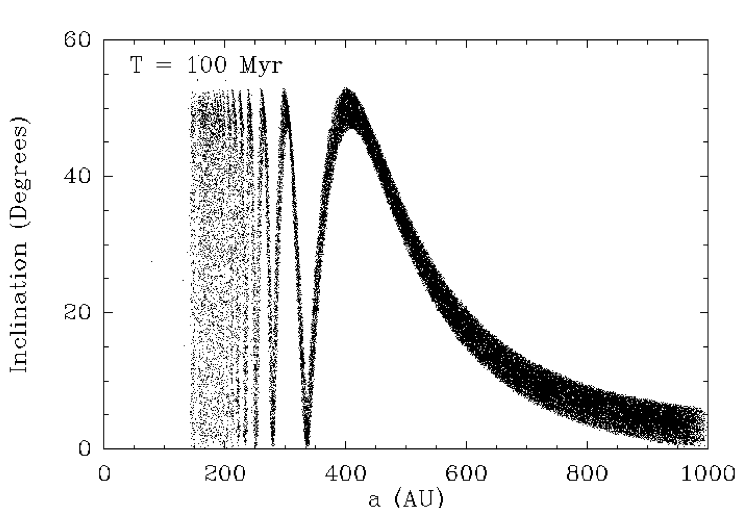}%
  \label{fig:sfig2_avsi}%
}
\caption{The left image shows an edge-on density map of a debris disk going from 143.1 to 1000\,AU after having an $25\deg$ inclined $1~M_j$ planet acting on it for 100\,Myr. The right image is a semimajor axis vs. particle inclination plot for the same system.}
\label{fig:1m25i_far}
\end{figure*}

We display the results of a simulation with a 1000\,AU disk over 100\,Myr in Figure~\ref{fig:1m25i_far}, for the planet with $M_p=1 M_\mathrm{J}$.  The disk develops a warped structure in which the innermost regions have enhanced inclinations, while the outermost parts still possess their initial inclinations.  The inclination distribution appears uniform within about 250\,AU, where $t_{sec}\,\approx\,30$\,Myr.  It thus appears that particles develop this uniform inclination distribution on the order of a few local $t_{sec}$.  After a few local secular timescales, the inclinations between nearby particles are no longer correlated and the disk appears inclined at that semi-major axis.  The disk is only completely inclined once the particles at the outermost edge undergo this interaction.

Finally, while \citet{Faramaz2014} found that the global eccentricities of disks acted upon by eccentric planets could relax in some cases, we found that the disks seemed to retain their heightened global inclination dispersions on Gyr timescales.

\subsection{Density Structure}
\label{ssec:density}

\begin{figure}
\centering
\includegraphics[width=1.\linewidth]{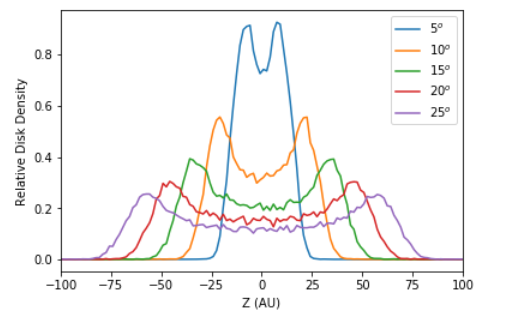}%
\caption{A plot of Z (vertical distance from the disk midplane) versus relative average column density, as seen in our $1~M_j$ simulations.}
    \label{fig:density_structure}
\end{figure}

\cite{Murray1999} states that particles in a disk perturbed by an inclined object will precess about a mean plane determined by the inclination of the planet relative to the disk.  As this motion is simple harmonic, the particles will tend to apparently ``clump" in regions furthest from the midplane.  This can be seen in the density map of the final disk structure in Figure~\ref{fig:density_structure}, in which we plot the column density (averaged over major axis of the disk when viewed edge-on) of the debris disk, viewed edge-on, versus the distance $Z$ from the midplane, in AU.  We see that, for all simulated $I_p$, the particle density has a bimodal distribution with a local minimum in the midplane.

Using purely geometrical arguments, we expect the location of the local maxima in density -- defined by its semimajor axis $a_\mathrm{max}$ within the disk midplane, and vertical distance from the midplane $Z_\mathrm{max}$ -- to be related to the initial planet misalignment $I_p$ as follows:
\begin{equation}
\label{eqn:zmax}
Z_\mathrm{max} \approx  \pm a_\mathrm{max} \times \mathrm{tan}(I_p) \qquad.
\end{equation}

This equation only applies if the eccentricity of the disk is zero, as the argument of periastron has an influence on the observed vertical disk width.  If $\omega_p = 0\degr$, the disk is inclined upon its minor axis, which would make it symmetric when viewed edge-on, meaning that the apocenter and pericenter will be similar in terms of vertical disk extent.  Meanwhile, if $\omega_p = 90\degr$ the disk will have a larger vertical extent around the apocenter than around the pericenter when viewed edge-on.  In these simulations, we have let $\omega_p = 0$, which means that we can replace $a$ in the previous equation with the semi-minor axis, $b = a\sqrt{1-e^2}$.  If we had instead allowed $\omega_p=90^o$, the disk would be asymmetric, replacing $a$ with $a(1-e)$ around the pericenter and with $a(1+e)$ around the pericenter.  The vertical distance to the midplane would vary by $2ae \mathrm{tan}I_p$ across the disk.  These two stated values of $\omega$ merely represent edge cases.  All other disks will possess some intermediate degree of asymmetry.  Equation~\ref{eqn:zmax} also assumes that the initial inclination dispersion of the disk is negligible compared to $I_p$.  There will be some dispersion in $Z_\mathrm{max}$ based upon the initial inclination dispersion of the disk particles.






Additionally, Figure~\ref{fig:density_structure} shows that the disk of a high-$I_p$ planet is more diffuse than a disk with a similar particle mass but a low-$I_p$ planet.  This would result in the high-$I_p$ disk having a lower maximum column density of particles when viewed edge-on.  This effect would be far less dramatic in systems that are face-on or close to face-on.  This is due to the fact that, in face-on systems, the vertical extent of the disk is difficult (if not impossible) to study.

\section{Application to the disk of HD 202628} \label{sec:disc}
In Section~\ref{ssec:density}, we found that an inclined planet would leave a clear footprint on the disk of HD 202628, and that this footprint is the most dramatic when viewed edge-on and is most difficult to observe face-on.

A planet that is inclined relative to the initially-flat debris disk shapes the disk into a spherical distribution of particles with an angular extent that goes roughly from $0\,$--$\,2\,I_p$. When viewed from large system inclinations, this can have a dramatic impact on the observational characteristics of HD 202628.  As the height of this spherical section is dictated by the planet's initial mutual inclination with the disk (as per Equation~\ref{eqn:zmax} and modified by the planet's eccentricity), both the apparent debris disk width and the way its intensity varies across the disk could theoretically be used to constrain $I_p$ from observations. The impact of a vertical disk extent has been studied in other inclined disks, such as HD 181327 \citep{Marino2016}, HR 4796A \citep{Kennedy2018}, and q$^1$ Eri \citep{Lovell2021}.

For the purposes of comparing our N-body simulations to the ALMA observations of HD 202628, we simulated observations of our inclined models.  To do so, we first took the 3D particle maps output by SWIFT-RMVS and rotated them such that they would match the observed $i_d$ and $\Omega_d$ observed by ALMA.  We then converted these into 2D density maps, which we used as an input into the ALMA Observation Support Tool\footnote{https://almaost.jb.man.ac.uk/}, assuming that the 1.3 mm observations would effectively trace the parent body distribution of the disk.  The antenna configuration and declination of observations was set to match the observations, and we varied the exposure time to match the observed sensitivity.  We normalized our inputs such that the total disk flux was 959\,$\mu$Jy.  We did not consider radiative transfer in these simulated images, as, at ALMA wavelengths, the azimuthal density variations of the particles is the dominant source of the disk flux variations.

\begin{figure*}
\centering
\includegraphics[width=0.3\linewidth]{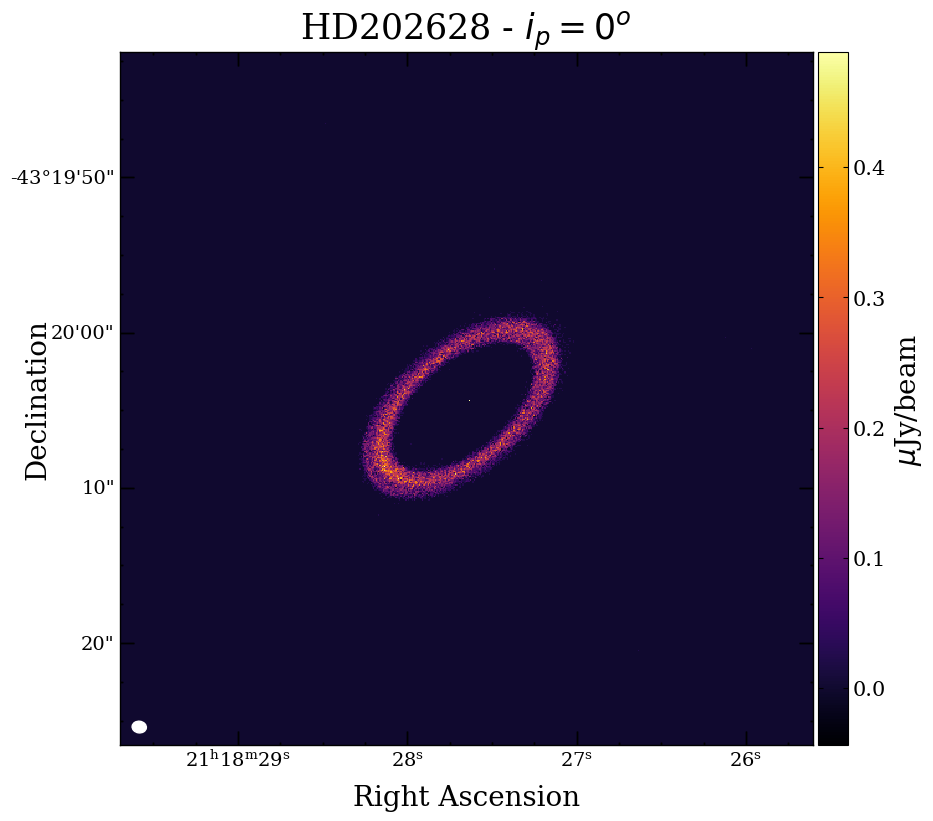}
\includegraphics[width=0.3\linewidth]{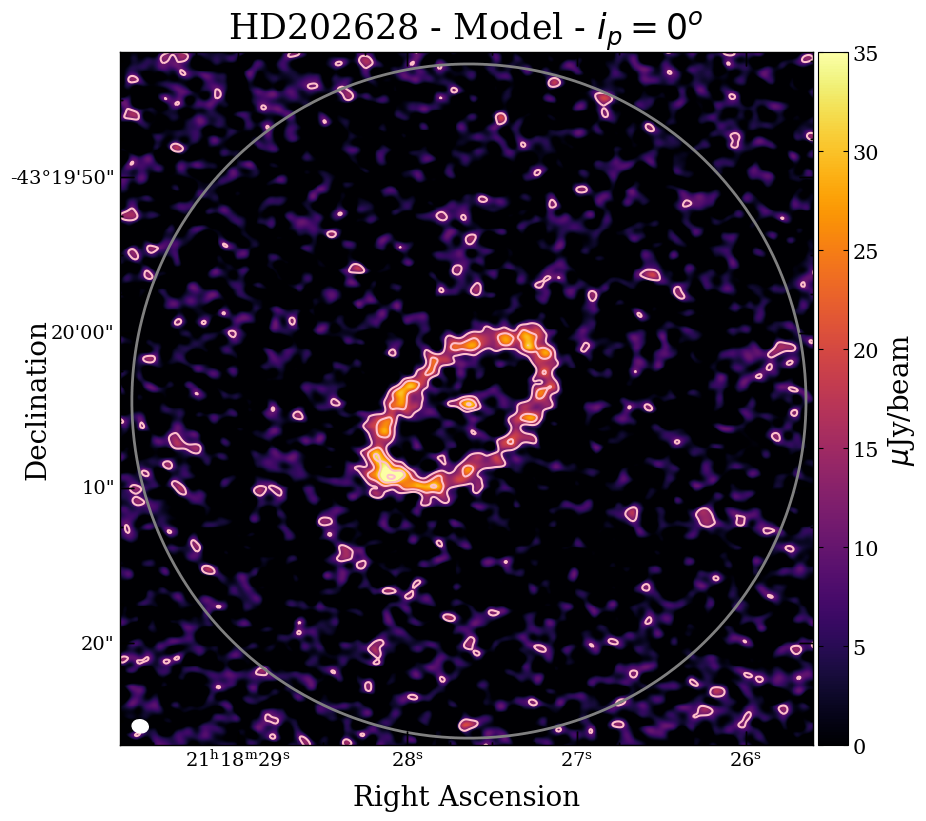}
\includegraphics[width=0.3\linewidth]{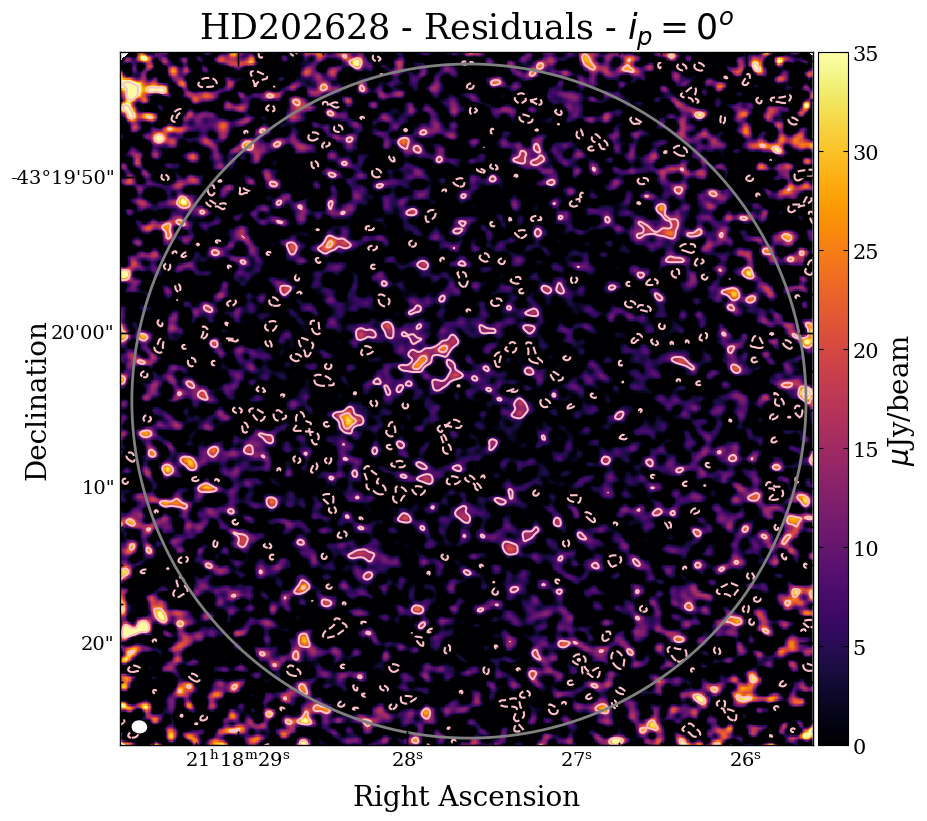}

\includegraphics[width=0.3\linewidth]{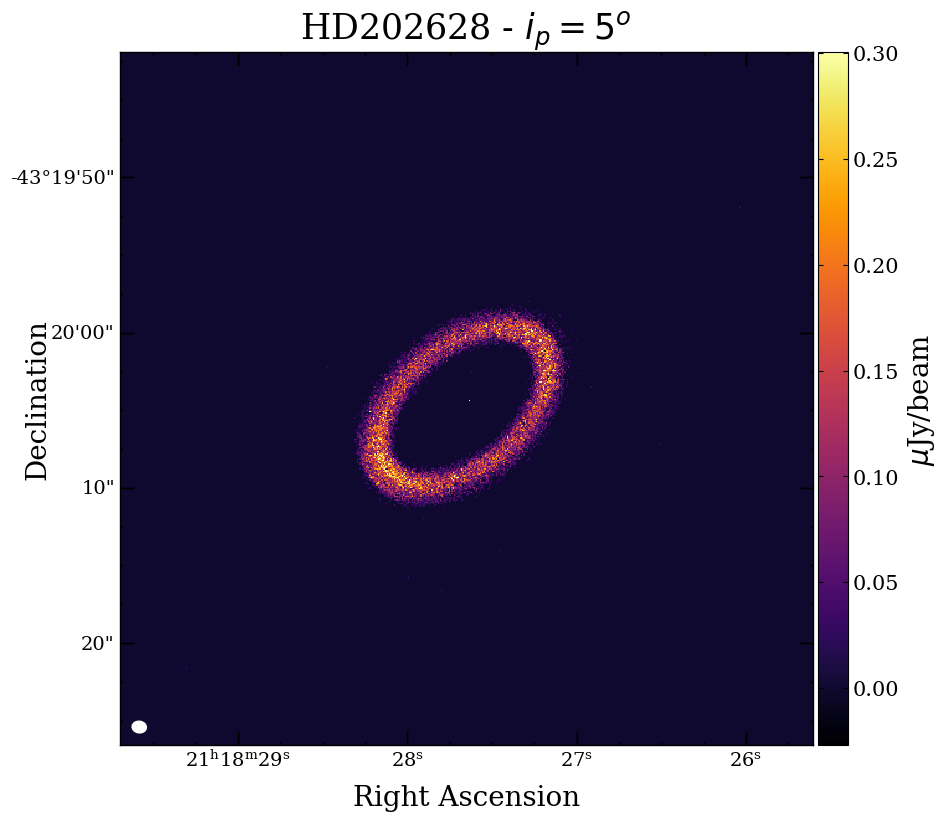}
\includegraphics[width=0.3\linewidth]{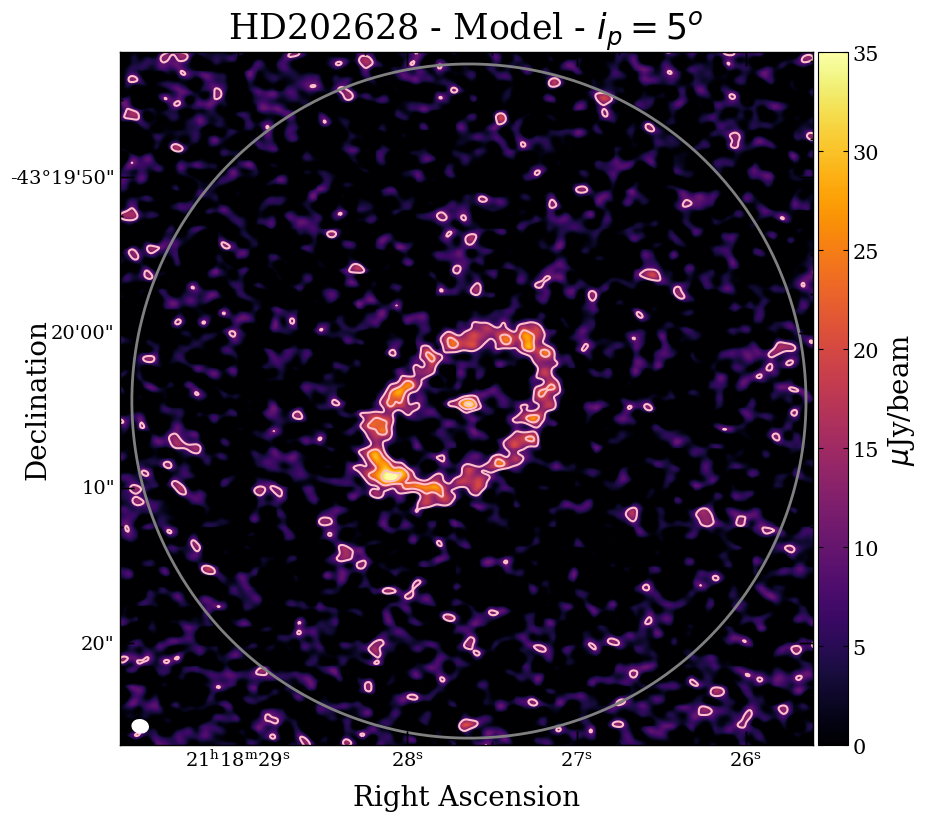}
\includegraphics[width=0.3\linewidth]{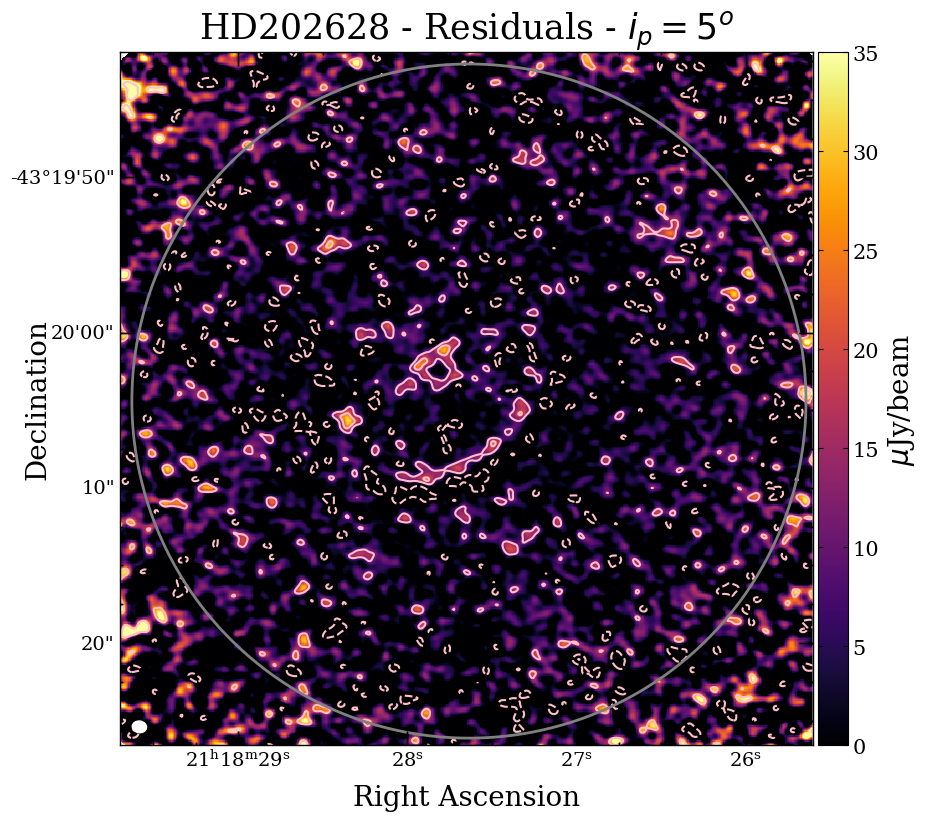}

\includegraphics[width=0.3\linewidth]{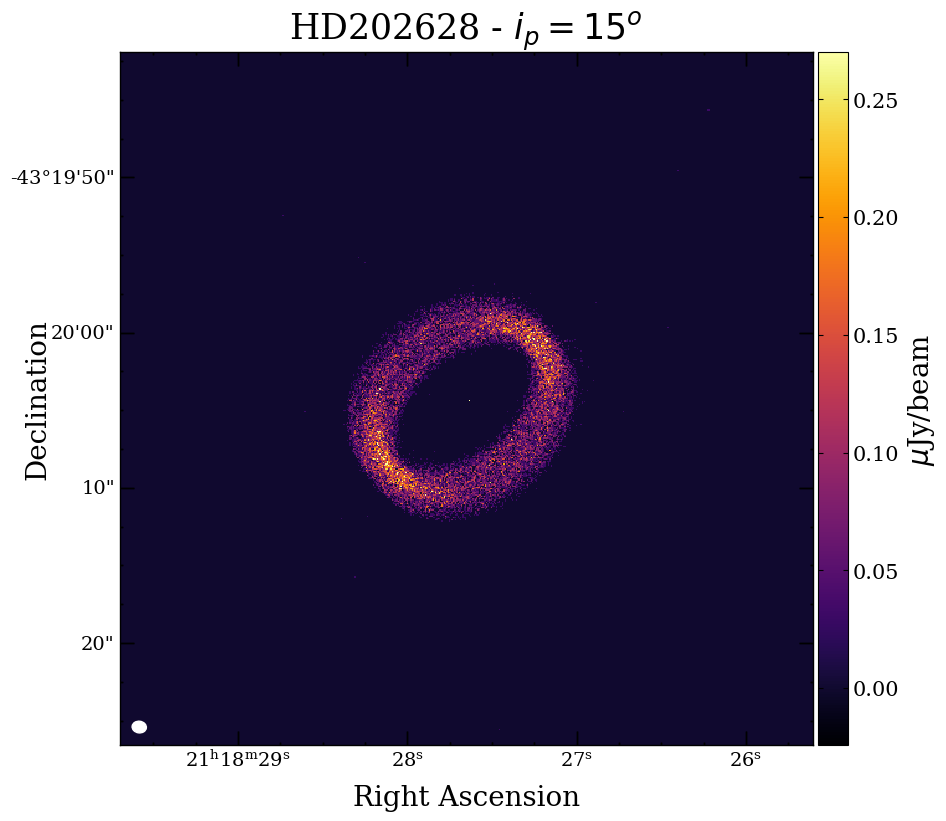}
\includegraphics[width=0.3\linewidth]{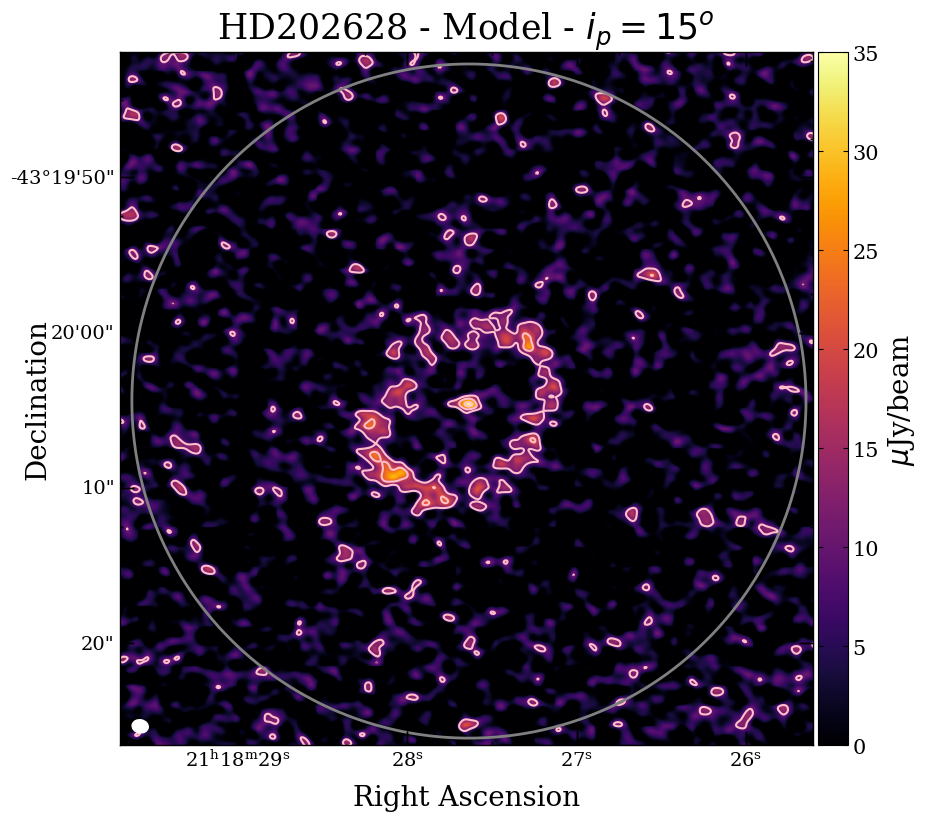}
\includegraphics[width=0.3\linewidth]{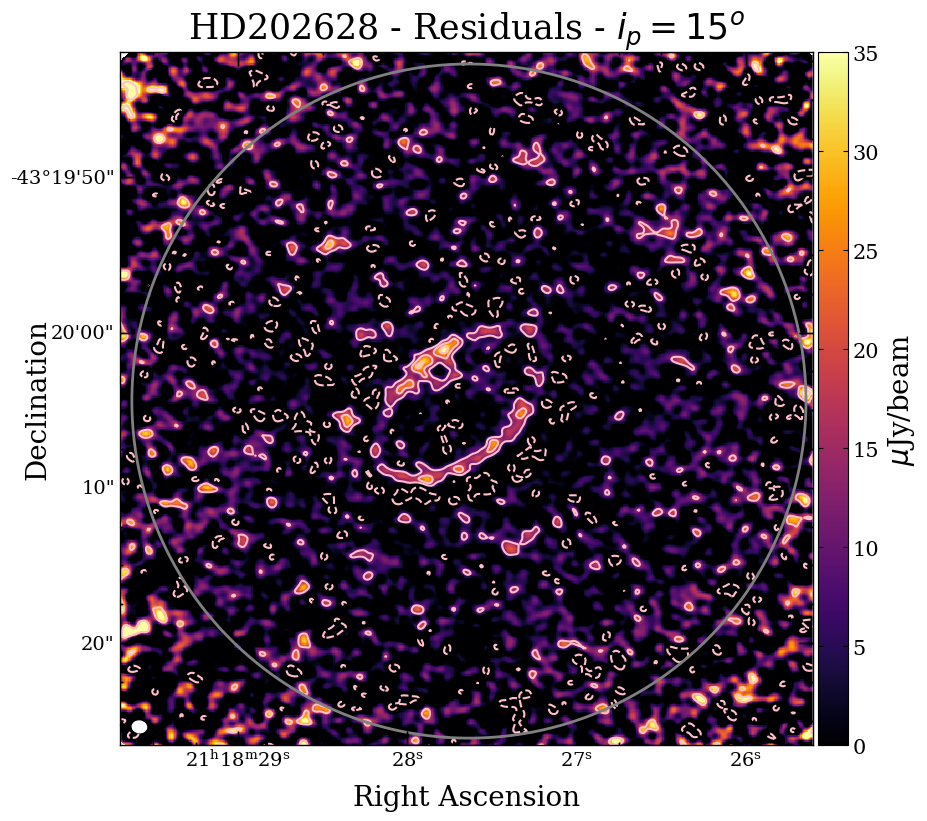}

\includegraphics[width=0.30\linewidth]{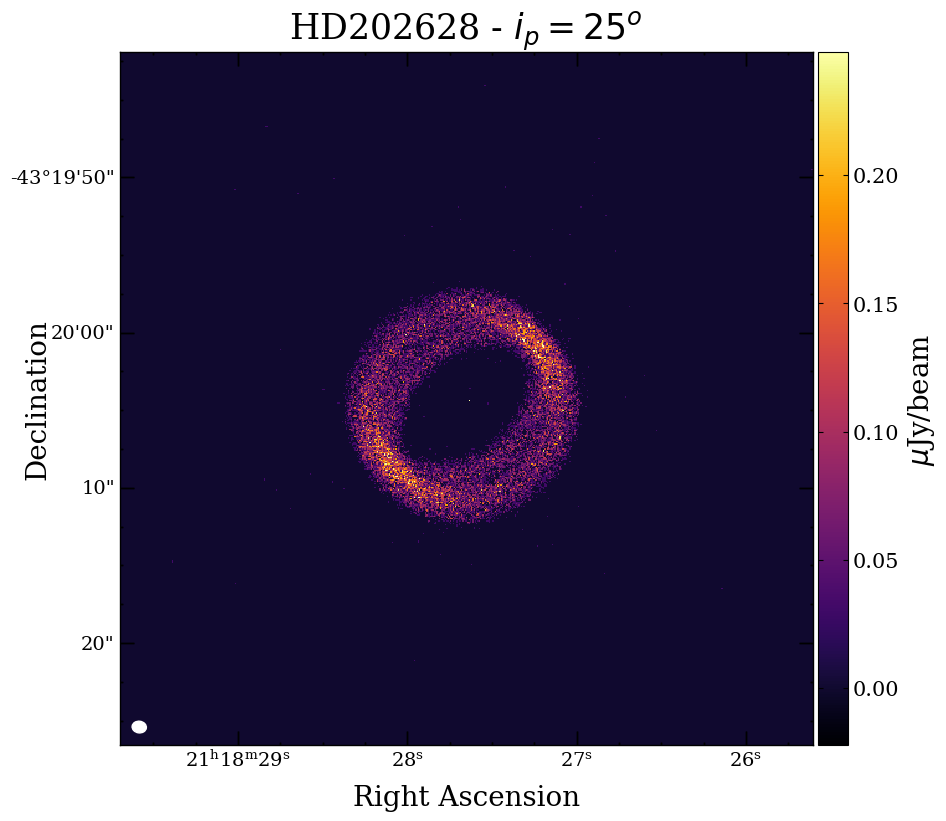}
\includegraphics[width=0.30\linewidth]{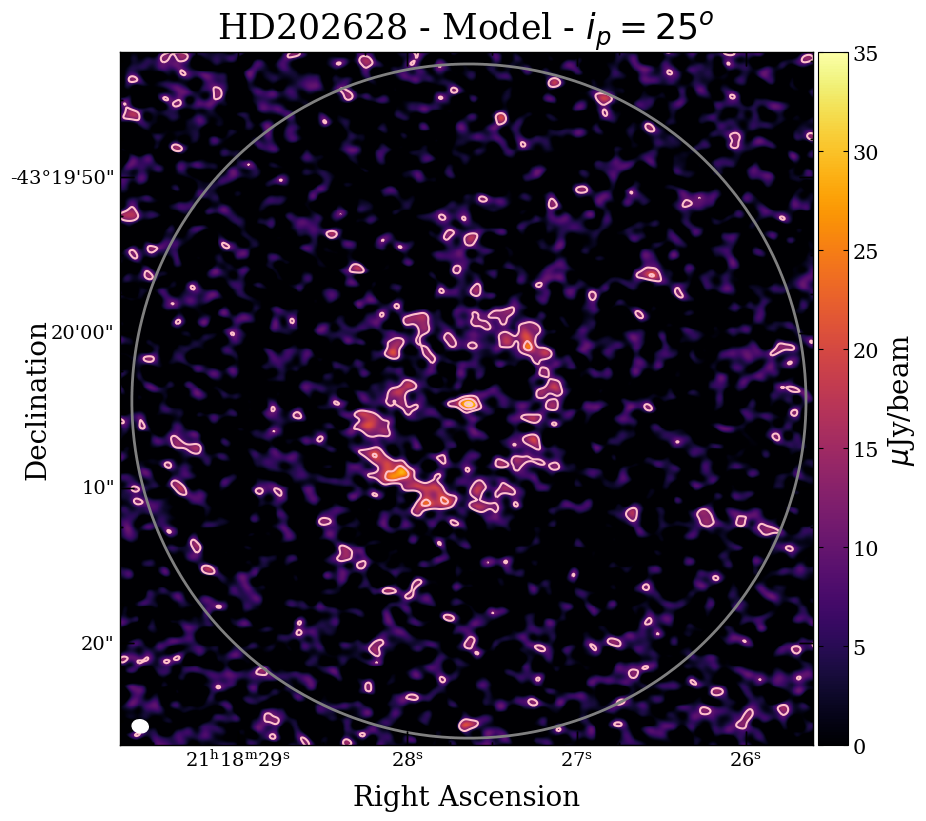}
\includegraphics[width=0.30\linewidth]{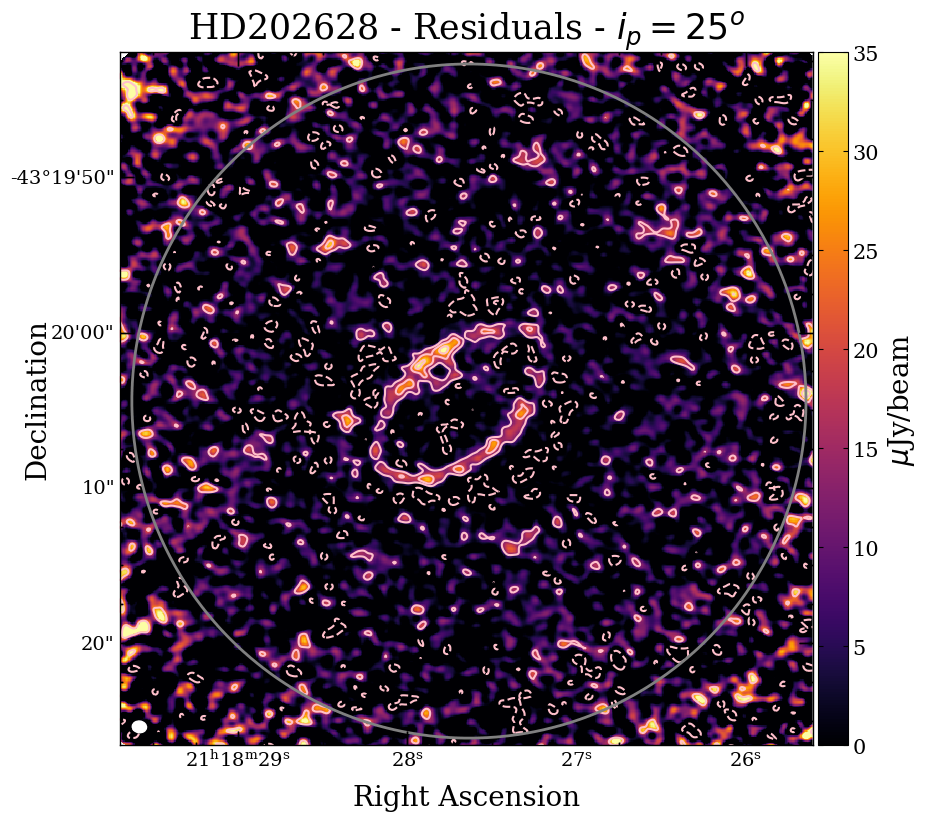}

\caption{Density maps of the $I_p = 0\deg,~5\deg,~15\deg$ and $25\deg$ models \emph{(left panels, top to bottom)}, simulated observations \emph{(center panels)}, and corresponding observational residuals \emph{(right planels)} using CASA. Fluxes and angular resolutions are match the ALMA observations. Contours are drawn at $2\sigma, 4\sigma, ...~\mu$Jy and $..., -4\sigma, -2\sigma, 2\sigma, 4\sigma, ...~\mu$Jy, for the simulated images and residuals, respectively, and with $\sigma = 5.2\,\mu$Jy/beam.}
    \label{fig:sim_obs1}
\end{figure*}

Figure~\ref{fig:sim_obs1} shows the density distributions for select models, along with simulated observations and residuals when compared to the ALMA observations. As each image is normalized for disk flux, the puffier, more diffuse disks (associated with planets with large $I_p$) have a lower surface brightness due to their larger area on-sky, and are thus more difficult to detect.  In addition, despite the fact that every system has the same eccentricity and is being viewed from the same angle, models with increasing $I_p$ appear more circular, as the disk particle distribution becomes more spherical, though this does not affect the location of the center offset.  The disk width is maximized and surface brightness is minimized on the sides pointing directly towards and away from us.  This effect would be more dramatic if we were viewing it at a system inclination higher than that referenced in Table~\ref{tab:star}.

As shown in Figure~\ref{fig:sim_obs1}, the vertical density distribution is not radially symmetric despite the fact that $\omega_p\,=\,0$, with a higher density of material near the apocenter.  This is likely due to the fact that the forced particle eccentricities decrease with semi-major axis as a consequence of Equation~\ref{eqn:ef}, causing an apocenter glow effect.  The effect is weak, however, agreeing with the observations from \cite{Faramaz2019}, which only show an apocenter glow effect at a low significance.  The apocenter glow and the influence of a mutually inclined planet may need to be considered in future works when modeling the intensity distributions of more sensitive ALMA images of puffy or eccentric disks.


The right column of Figure~\ref{fig:sim_obs1} shows the residual plots between our models and the HD 202628 ALMA data.  It is obvious that the ALMA data does not support the presence of an exoplanet with a high mutual inclination relative to the initial disk.  An excited planetary inclination would result in intensity maxima (and width minima) across the major axis and intensity minima (and width maxima) along the minor axis.  However, both of these features in our observations can be explained with purely geometric effects and possibly other phenomena (such as apocenter glow).  The residual plots for all except the $I_p\,=\,0\deg$ model show clear traces of the disk, indicating that the true HD 202628 disk is consistent with a low-inclination planet.  Given our somewhat coarse grid of models, we thus conclude that any present planet likely has $I_p$ between $0\,$--$\,5\deg$ and exclude any $I_p$ greater than $I_p\,=\,5\deg$.  This is more constraining from what was found in \citet{Kennedy2020}, who found that $I_p<29\degr$.  One potential cause for this discrepancy is our usage of full N-body simulations as opposed to analytical methods, which potentially make use of simplifications.  Another potential difference is in our choices of parameters to vary.  As an example, \citet{Kennedy2020} found a correlation between the fit inclination and the semi-major axis of the disk, while we assumed a fixed semi-major axis.  However, we found that a planet inclination $I_p\,>\,10\degr$ causes a visually obvious variation in apparent disk width across the extent of the disk, which does not match our observations and cannot be explained by an error in semi-major axis fitting.  We thus conclude that differences between the analytical models and the N-body simulations are likely the cause of the different constraints.



As the system is a Gyr old, if there was a planet in the system with an excited inclination, we expect that it would have already completely inclined the disk at the time of observation unless it had a very low mass.  However, past studies by \cite{Thilliez2016} and \cite{Faramaz2019} have excluded such planets as possibilities.  If a planet existed in this system long enough to fully excite the disk eccentricity, it would have had enough time to incline the disk as well, as the two processes evolve along the same timescale.  This would be the case unless some other process causes significant damping of particle inclination or the eccentricity and inclination are influenced by two different planets, as per \citet{Dong2020}.  We can thus conclude that, if there is a single planet responsible for shaping the debris disk, it initially had a low mutual inclination with the disk.

We can study the results from \cite{Carrera2019} to determine what a low mutual inclination can tell us about the evolution history of the disk-stirring planet.  Their simulations show that scattering events seem to frequently excite the planets' inclinations, but as many as 30\% of the $a_p\,>\,50$\,AU planets in their simulations had final inclinations $<\,5\deg$.  Thus, the low observed inclination of the HD 202628 disk ($I_p\,<\,5\deg$) does not eliminate the possibility that the planet's long-period orbit and excited eccentricity were the result of scattering events.

\section{Conclusion} \label{sec:conclu}
Exoplanets can completely incline debris disks over timescales similar to those at which they excite disk eccentricities \citep[see, e.g.,][]{Pearce2014}.  These timescales are highly mass-dependent and are small compared to the system age of HD 202628, meaning that we expect to observe the signature of any inclined planet on the disk, especially since we expect this puffed structure to be preserved throughout the system's lifetime.  

Due to the evolution of particle inclinations with time, the modeled disk developed a density structure with a height dictated by the disk's extent, the initial mutual inclination of the planet with the disk, and the planet's eccentricity.  This structure results in a concentration of particles at high Z. Viewing the system in any orientation that is not face-on allows us to observe a puffy debris disk with a variable width due to both geometric effects and the density structure.

As we are not viewing HD 202628 face-on, these characteristics are observable and can be used to constrain the mutual inclination of a planet in the disk, as an inclined planet can ``puff up" a disk in an observable fashion.  If not accounted for in simulations, this could lead to an incorrect determination of the disk properties, such as eccentricity, disk width, or semi-major axis.  As an example, the warp seen in scattered light with the SPHERE instrument at the VLT in the edge-on debris disk of HD 110058 \citep{Kasper2015} and its unusually large vertical extent seen at ALMA wavelengths \citep{Hales2022}, suggests the presence of a planet on an orbit significantly inclined compared to the disk (Stasevic et al., in prep) and might justly illustrate the findings of our study and be used to test and confirm them.

However, in the case of the system of HD 202628, our simulations show that any potential stirring planet would have a mutual inclination of less than $5\deg$ with the initial disk. While low, this inclination does not exclude the possibility that the planet is on its current orbit as the result of scattering events.  Higher SNR observations or a finer grid of model simulations could provide more precise insights into the exact nature of this disk and its potential dynamical history.

Finally, the results of our work validate the assumption of disk-planet coplanarity used in dynamical studies that focused on debris disks that appear as narrow rings -- such as HR 4796 \citep{Wyatt1999,Lagrange2012,Kennedy2018}, or Fomalhaut \citep{Quillen2006,Chiang2009,Boley2012,Beust2014,Faramaz2015,Pearce2021}.

\begin{acknowledgments}
This research has made use of the NASA Exoplanet Archive, which is operated by the California Institute of Technology, under contract with the National Aeronautics and Space Administration under the Exoplanet Exploration Program.  This material is based upon work supported by the National Science Foundation Graduate Research Fellowship under Grant No.\ DGE 1746045.  This research has made use of NASA's Astrophysics Data System Bibliographic Services. VF and SE acknowledge funding from the National Aeronautics and Space Administration through the Exoplanet Research Program under Grant No. 80NSSC21K0394 (PI: S. Ertel).
\end{acknowledgments}

%

\vspace{5mm}
\facility{Exoplanet Archive}







\bibliography{HD202628}{}
\bibliographystyle{aasjournal}



\end{document}